\documentclass[10pt,a4paper]{article}
\usepackage{jheppub_kim}
\usepackage{pdflscape}
\usepackage{amsmath}
\usepackage{amssymb}
\usepackage{dcolumn}
\usepackage{bm}
\usepackage{color}
\usepackage{epsfig}
\usepackage{amsfonts}
\usepackage{graphicx}
\usepackage{subfigure}
\usepackage{dcolumn}
\UseRawInputEncoding

\begin{document}
\title{Gravitational Collapse in Energy-momentum squared gravity: Nature of singularities}

\author[a]{Prabir Rudra}

\affiliation[a] {Department of Mathematics, Asutosh College,
Kolkata-700 026, India}

\emailAdd{prudra.math@gmail.com, prabir.rudra@asutoshcollege.in}

\abstract{In this paper we explore a collapsing scenario in the
background of energy-momentum squared gravity (EMSG). EMSG claims
to have terms that originate from the quantum gravity effects
mimicking loop quantum gravity. As a result the framework admits a
bounce at a finite time thus avoiding a singularity. So the
question that naturally arises : Is there any realistic chance of
formation of a black hole or the quantum gravity effects are
strong enough to totally avoid such a pathology? Motivated from
this we are interested in studying a gravitational collapse
mechanism in the background of EMSG and investigate the fate of
such a process. We model the spacetime of a massive star by the
Vaidya metric and derive the field equations in EMSG. Then using
the equations we go on to study a gravitational collapse
mechanism, on two specific models of EMSG with different forms of
curvature-matter coupling. The prime objective is to probe the
nature of singularity (if formed) as the end state of the
collapse. We see that none of the models generically admit the
formation of black holes as the end state of collapse, but on the
contrary they support the formation of naked singularities. This
can be attributed to the quantum fluctuations of the gravitational
interactions at the fundamental level.}


\maketitle

%%%%%%%%%%%%%%%%%%%%%%%%
\section{Introduction}
%%%%%%%%%%%%%%%%%%%%%%%%
The incompatibility of General Relativity (GR) at cosmological
scales came to light with the discovery of the late cosmic
acceleration \cite{acc1, acc2}. This triggered extensive
theoretical research to develop a framework that can
comprehensively explain the phenomenon. Over the past two decades
researchers have proposed two independent mechanism that can
satisfactorily explain the cosmic acceleration, namely the theory
of dark energy \cite{de1} and the theory of modified gravity
\cite{mod1, mod2, mod3}. The first one proposes the presence of an
exotic fluid with negative pressure that can drive the
acceleration. The second concept is based on the modification of
the Einstein-Hilbert (EH) action of GR, where the the gravity
Lagrangian (the Ricci scalar $R$) is replaced by some other terms.
In this work we will concentrate on this second avenue of research
that has gained enormous success and popularity is the past
decades.

When we talk about modification of the Einstein gravity, the most
straightforward modification of the gravity Lagrangian will be to
consider an analytical function of the Ricci scalar $f(R)$, which
will help explore the non-linear effects of the curvature (may be
the most fundamental driving force behind the acceleration). Such
theories are termed as $f(R)$ gravity theories in literature
\cite{frorg, fr1, fr2}. Once such a modification was found to be
consistent with the theoretical framework of GR and gave promising
results, it was inevitable that other factors were included in the
functional forms to extend the theory. Such a crucial extension
was proposed by Harko \& Lobo in ref.\cite{frlm}, where they
considered an even more general functional form in the
gravitational Lagrangian by coupling the matter Lagrangian
$\mathcal{L}_{m}$ with $R$. Such theories are known as
$f(R,\mathcal{L}_{m})$ theories and significant advances in this
area may be found in refs.\cite{frlm2, frlm3, frlm4}. Following
similar concept Harko et al. \cite{harko1} proposed the $f(R,T)$
theory where they considered the trace of the energy-momentum
tensor $T$ in place of the matter Lagrangian in the
$f(R,\mathcal{L}_{m})$ theory. Here due to the coupling effects
between matter and geometry sectors, the covariant divergence of
the energy-momentum tensor (EMT) is non-zero. This indicates
non-convergence of EMT leading to extra force, that results in
non-geodesic nature of motion of the particles. This interesting
feature of the theory attracted a lot of research on it leading to
substantial developments \cite{frt1, frt2, frt3, frt4, frt5, frt6,
frt7}. In a subsequent modification $f(R,\textbf{T}^{2})$ theory
was proposed in \cite{emsg1}, where
$\textbf{T}^{2}=T_{\mu\nu}T^{\mu\nu}$ and $T_{\mu\nu}$ is the EMT.
A specific form of $f(R,\textbf{T}^{2})=R+\eta \textbf{T}^2$, was
studied in \cite{emsgorg}, which was termed as energy-momentum
squared gravity (EMSG). The study revealed that the field
equations of EMSG contained terms similar to those arising from
the quantum gravity effects of loop quantum gravity (LQG) and
brane gravity. The theory allowed for a minimum length and a
maximum energy density in the early universe. In a Friedmann
universe it allows for a cosmic bounce alleviating the singularity
problem. Subsequent developments in EMSG can be found in
literature. Cosmological models in EMSG was studied in
\cite{board1, akarsu1}. A dynamical system analysis in the
background of EMSG was performed in \cite{rudra1}. Other important
developments in EMSG may be found in \cite{rudra2, rudra3,
akarsu2, akarsu3, moraes1, nari1, keskin1}.

A star at the end of its lifetime exhausts all its nuclear sources
of energy and undergoes gravitational collapse due to its own
gravitational field and releases enormous amount of energy.
Gravitational collapse is a crucial astrophysical phenomenon that
is responsible for the structure formation in the universe.
Various stellar remnants like the white dwarfs, neutron stars,
black holes, etc. are the end products of gravitational collapse
of stars. The type of end-product that a gravitationally
collapsing system will produce depends on the initial mass
configuration of the parent stellar system. A small or an average
sized star may end its collapsing process in a white dwarf, where
further collapse is hindered by the electron degeneracy pressure.
But subsequently if the white dwarf accretes enough mass from the
surrounding so that it surpasses the Chandrasekhar limit ($> 1.39$
solar mass), then the collapse resumes and subsequently it reaches
the neutron star stage. In this stage the further collapse is
hindered by the neutron degeneracy pressure, which is overcome by
again accreting mass from the surrounding. Then further collapse
begins with a simultaneous supernova event, and this time it ends
in a singularity. So it is clear that throughout the whole
collapsing sequence, the most important role is played by mass of
the star. The collapse of a super-massive star, having a huge
initial mass ($>30$ solar masses) will seldom be terminated at any
of the intermediate stage (white dwarf or neutron star), but will
eventually reach a singularity. The astrophysical significance of
the gravitational collapse attracts a lot of exploration on the
topic. It all started in $1939$ with the pioneering work of
Oppenheimer and Snyder \cite{oppen}, studying the collapse of a
dust cloud modelled by a Schwarzschild exterior and a Friedmann
interior. This was followed by Tolman \cite{tolman} and Bondi
\cite{bondi} who studied the collapse of an inhomogeneous
distribution of dust with a spherically symmetric distribution.
All these collapsing scenarios resulted in the formation of black
holes. Extensive reviews in gravitational collapse may be found in
\cite{collrev1, collrev2}. In 1969 Roger Penrose proposed his
famous {\it Cosmic Censorship Hypothesis (CCH)} \cite{penrose},
which stated that any singularity formed out of a gravitational
collapse will always be a trapped inside an event horizon (black
hole). This meant a complete censorship of the singularity from
the outside observers such that no information of the interior is
leaked to the outside. This phenomenon of black holes is termed as
the information loss paradox \cite{paradox1, paradox2, paradox3,
paradox4}. But due to the absence of a comprehensive proof of the
CCH, there was widespread doubts regarding its validity.
Scientists began searching for ways that can disprove the
hypothesis. Now in order to accomplish that, it was necessary to
show that a collapsing process does not always result in a trapped
surface, but also has provisions for the formation of a {\it naked
singularity (NS)} \cite{ns1, ns2, ns6, ns7, lake2, szek, ghosh1}.
For such a singularity, no event horizon forms and there can be a
continuous exchange of information to and from the singularity.
This directly solves the information loss problem. Not only this,
with the continuous supply of data right from the heart of the
singularity, our quantum gravity problem will gain impetus. As a
result a hunt for NS began with the works of Eardly and Smarr
\cite{ns1} and Christodoulou \cite{ns2} at the end of the
twentieth century. Other significant works in NS can be widely
found in literature in the refs.\cite{ns3, ns4, ns5}.

The AdS/CFT correspondence \cite{ads1, ads2} has been used to gain
insights into the behavior of strongly interacting quantum
systems, including those that might be analogous to certain
aspects of black hole physics. In the context of gravitational
collapse, researchers have explored the idea that the dynamics of
a collapsing star and the resulting black hole formation might be
dual to a description in terms of a non-gravitational quantum
field theory living on the boundary of an AdS spacetime. This
holographic duality allows physicists to study certain
gravitational phenomena, such as black hole formation and
evaporation, through the lens of a non-gravitational theory. It
provides a different perspective on understanding the quantum
nature of gravity in extreme conditions, where classical general
relativity alone may not be sufficient. Moreover extensive use of
the holographic principle can be found in cosmology \cite{hp1,
hp2}. It's important to note that the application of holography to
gravitational collapse and cosmology is an active area of
research, and the full understanding of these connections is still
evolving.

Schwarzschild metric represented the spacetime outside a
spherically symmetric matter distribution possessing a constant
mass. Such a solution is static in nature and cannot model the
spacetime of a realistic star. A solution to this problem was
given by Vaidya in his ground breaking paper in 1951
\cite{vaidya}, where he proposed a dynamic spacetime outside a
realistic star. Vaidya's metric was aptly termed as the {\it
shining or radiating Schwarzschild metric.} Here the constant mass
parameter in the Schwarzschild metric was replaced by a time
dependent dynamic mass parameter, which was the source for the
outgoing stellar radiations. Various studied in Vaidya spacetime
can be found in the refs.\cite{s1, s2, s3, s4, s5, s6, s7, s8, s9,
s10}.

In this work we intend to study a stellar collapse modelled by
Vaidya spacetime in the background of energy-momentum squared
gravity. Since EMSG is known to contain some flavours of quantum
gravity we are motivated to explore a collapsing scenario in this
gravity, that will continue the whole way till the singularity
(collapse of a super-massive star). Our basic plan is to study the
nature of singularity (BH or NS) that results from the collapse
(if at all a singularity forms). This will throw new light on the
terms that provide the quantum gravitational effects in EMSG and
also the validity of CCH. The fact that GR foretells the
occurrence of spacetime singularity at some finite point in the
past is one of the theory's most fascinating mysteries. However,
because of the anticipated quantum consequences, it turns out that
GR itself is no longer valid at the singularity. However, there is
currently no accurate description of quantum gravity. When we
apply this theory to a homogeneous and isotropic spacetime, we
discover that the early Universe has a minimum length scale and a
maximum energy density. In other words, there is an early-time
bounce and, as a result, the early-time singularity is avoided.
This is the main consequence of the correction terms that appear
in the theory. This is the reason why these terms are compared to
those of the Loop quantum gravity and brane-world gravity. While
the underlying physical theory does not incorporate the LQG or
brane-world models directly, nor does it reduce to them in a
limiting case, it focuses on a type of higher-order matter
corrections that modify the Friedmann equations in ways that
include phenomenological modifications coming from both Loop
quantum gravity and brane-world gravity. It makes sense to assume
that this correction term would only matter in high-energy
environments, such the early Universe or black holes.
Consequently, in the low-curvature regime, there are no deviations
from GR. This is the reason why there is a strong motivation of
studying the nature of singularities arising from gravitational
collapse in the background of this theory. The quantum
gravitational effects coming from the correction terms of this
theory has crucial physical significance and interpretation in the
study of the fate of these singularities.

The paper is organized as follows: Section 2 is dedicated in
deriving the field equations of Vaidya spacetime in
$f(R,\mathbf{T}^2)$ gravity. In section 3 we explore a collapsing
mechanism using the derived field equations. Finally the paper
ends with a discussion and conclusion in section 4.

%%%%%%%%%%%%%%%%%%%%%%%%%%%%%%%%%%%%%%%%%%%%%%%%
\section{Vaidya spacetime in $f(R,\mathbf{T}^2)$ gravity}
%%%%%%%%%%%%%%%%%%%%%%%%%%%%%%%%%%%%%%%%%%%%%%%%
The action of the model is written as~\cite{emsg1, board1}
\begin{equation}
\mathcal{S}=\frac{1}{2\kappa^2}\int d^4x \sqrt{-g}
f(R,\mathbf{T^2}) + \int d^4x \sqrt{-g} \mathcal{L}_{m},
\end{equation}
where $f$ is a function depending on the square of the
energy-momentum tensor $\mathbf{T^2}=T^{\mu\nu}T_{\mu\nu}$ and the
scalar curvature $R$. Here, $\kappa^2=8\pi G$ and
$\mathcal{L}_{m}$ represents the matter Lagrangian.

If we vary the action with respect to the metric we arrive at the
following field equations
\begin{equation}
R_{\mu \nu}f_R  +g_{\mu\nu} \Box f_R-\nabla_{\mu}\nabla_{\nu}
f_R-\frac{1}{2} g_{\mu\nu}f=\kappa^2 T_{\mu \nu}-f_{\mathbf{T^2}}
\Theta_{\mu \nu}\,, \label{FieldEq}
\end{equation}
where $\Box=\nabla_\mu \nabla^\mu$, $f_{R}=\partial f/\partial R$,
$f_{\mathbf{T^2}}=\partial f/\partial \mathbf{T^2}$ and
\begin{equation}
    \Theta_{\mu\nu}=\frac{\delta (\mathbf{T^2})}{\delta g^{\mu\nu}}=
    \frac{\delta (T^{\alpha\beta}T_{\alpha\beta})}{\delta g^{\mu\nu}}=
    -2\mathcal{L}_{m}\Big(T_{\mu\nu}-\frac{1}{2}g_{\mu\nu}T\Big)-T\, T_{\mu\nu}
    +2T^{\alpha}_{\mu}T_{\nu\alpha}-4T^{\alpha\beta}\frac{\partial^2
    \mathcal{L}_{m}}{\partial g^{\mu\nu}\partial g^{\alpha\beta}}\,,\label{Theta}
\end{equation}
where $T$ is the trace of the energy-momentum tensor. By taking
covariant derivatives with respect to the field
equation~\eqref{FieldEq}, one finds the following conservation
equation
\begin{eqnarray}
\kappa^2\nabla^\mu T_{\mu\nu}=-\frac{1}{2}g_{\mu\nu}\nabla^\mu
f+\nabla^\mu(f_{\mathbf{T}^2}\Theta_{\mu\nu})\,.\label{conservation1}
\end{eqnarray}
As one can see from the above equation that in general, the
conservation equation does not hold for this theory. If one
chooses $f(R,\mathbf{T}^2)=2\alpha \log(\mathbf{T}^2)$, one gets
the same result reported in Akarsu et al. \cite{akarsu1}.

Here we consider perfect fluid and further, we assume
$\mathcal{L}_{m}=p$ which allows us to rewrite $\Theta_{\mu\nu}$
defined in eqn. \eqref{Theta} as a quantity which does not depend
on the function $f$, as given below \cite{emsg1, board1}
\begin{equation}
\Theta_{\mu\nu}=-\Big(\rho^2+4 p\rho+3p^2\Big)u_\mu u_\nu\,.
\end{equation}
Moreover
\begin{equation}
\mathbf{\Theta^2}:=\Theta_{\mu\nu}\Theta^{\mu\nu}=\rho^2+4p
\rho+3p^2\label{Theta2}
\end{equation}
is defined.

The Vaidya metric in the advanced time coordinate system is given
by \cite{vaidya},
\begin{equation}\label{vaidya}
ds^{2}=f(t,r)dt^{2}+2dtdr+r^{2}\left(d\theta^{2}+\sin^{2}\theta
d\phi^{2}\right)
\end{equation}
~~~~~where $f(t,r)=-\left(1-\frac{m(t,r)}{r}\right)$ and using the
units $G=c=1$. The total energy momentum tensor of the field
equation (\ref{FieldEq}) is given by the following sum,
\begin{equation}\label{energymom}
T_{\mu\nu}=T_{\mu\nu}^{(n)}+T_{\mu\nu}^{(m)}
\end{equation}
where $T_{\mu\nu}^{(n)}$ and $T_{\mu\nu}^{(m)}$ are the
contributions from the Vaidya null radiation and matter
respectively defined as,
\begin{equation}\label{null}
T_{\mu\nu}^{(n)}=\sigma l_{\mu}l_{\nu}
\end{equation}
and
\begin{equation}\label{fluid}
T_{\mu\nu}^{(m)}=(\rho+p)(l_{\mu}\eta_{\nu}+l_{\nu}\eta_{\mu})+pg_{\mu\nu}
\end{equation}
where $'\rho'$ and $'p'$ are the energy density and pressure for
matter and $'\sigma'$ is the energy density corresponding to
Vaidya null radiation. In the co-moving co-ordinates
($t,r,\theta_{1},\theta_{2},...,\theta_{n}$), the two eigen
vectors of energy-momentum tensor namely $l_{\mu}$ and
$\eta_{\mu}$ are linearly independent future pointing null vectors
having components
\begin{equation}\label{vectors1}
l_{\mu}=(1,0,0,0)~~~~ and~~~~
\eta_{\mu}=\left(\frac{1}{2}\left(1-\frac{m}{r}\right),-1,0,0\right)
\end{equation}
and they satisfy the relations
\begin{equation}\label{vectors2}
l_{\lambda}l^{\lambda}=\eta_{\lambda}\eta^{\lambda}=0,~
l_{\lambda}\eta^{\lambda}=-1
\end{equation}
Therefore, the non-vanishing components of the total
energy-momentum tensor will be as follows
\begin{eqnarray*}
T_{00}=\sigma+\rho\left(1-\frac{m(t,r)}{r}\right),&&
~~T_{01}=-\rho \\
\end{eqnarray*}
\begin{eqnarray}\label{energymomentum}
T_{22}=pr^2, && ~~T_{33}=pr^2 \sin^2\theta
\end{eqnarray}
Here we consider matter in the form of perfect barotropic fluid
given by the equation of state
\begin{equation}\label{pressure}
p=\omega\rho
\end{equation}
where '$\omega$' is the barotropic parameter.

The non-vanishing components of the Ricci tensors are given by,
\begin{eqnarray*}
R_{00}=\frac{\left(m-r\right)m''+2\dot{m}}{2r^{2}},&&
~~~~R_{01}=R_{10}=\frac{m''}{2r} \\
\end{eqnarray*}
\begin{eqnarray}\label{einsteintensors}
R_{22}=m', && ~~R_{33}=m'~ \sin^{2}\theta
\end{eqnarray}
where ~$.$~ and ~$'$~ represents the derivatives with respect to
time coordinate $'t'$ and radial coordinate $'r'$ respectively.
For this system the Ricci scalar becomes,
\begin{equation}\label{ricciscalar}
R=\frac{2m'+rm''}{r^{2}}
\end{equation}
The trace of the energy momentum tensor is calculated as,
\begin{equation}\label{trace}
T=g^{\mu\nu}T_{\mu\nu}=2\left(\omega-1\right)\rho
\end{equation}
The square of the energy momentum tensor will be given by,
\begin{equation}
\textbf{T}^2=T_{\mu\nu}T^{\mu\nu}=4p^{2}=4\omega^{2} \rho^{2}
\end{equation}

The relation between density and mass is considered as
\cite{boer1},
\begin{equation}\label{densitymass}
\rho=n\times m(t,r)
\end{equation}
where $n>0$ is the particle number density.

%%%%%%%%%%%%%%%%%%%%%%%%%%%%%%
\subsection{Field equations}
%%%%%%%%%%%%%%%%%%%%%%%%%%%%%%
From eqn.(\ref{FieldEq}), we see that we need to put specific
forms of $f(R,\mathbf{T}^2)$ to get further solutions. Here we
will derive the field equations of Vaidya spacetime in EMSG for
such specific models depending on the nature of coupling between
$R$ and $\mathbf{T}^2$. We basically consider two general forms,
one of which is additive representing minimal coupling and the
other is in the product form representing non-minimal coupling.
they are discussed below in Model-1 and Model-2 respectively.

\subsubsection{Model-1:  $f(R,\mathbf{T}^{2})=f_{1}(R)+f_{2}(\mathbf{T}^2)$}
Here we report the computed Einstein's field equations of
$f(R,\mathbf{T}^2)$ gravity in the time dependent Vaidya spacetime
for the first model.\\

\textbf{\textit{The (01) component is given by,}}
\begin{equation}\label{fieldeq01}
r\left(f_{1}(R)+f_{2}(T^{2})-2\rho\right)-f_{1}'(R)m''=0
\end{equation}

\textbf{\textit{The (22) and (33) components are given by,}}
\begin{equation}\label{fieldeq2233}
r^{2}\left(f_{1}(R)+f_{2}(T^{2})+2\omega\rho\right)-2f_{1}'(R)m'=0
\end{equation}

Here we have only reported those components which we have used to
find a solution. The other components of the field equations are
reported in the appendix section of the paper in order to preserve
a better structure of the paper.

\subsubsection{Model-2:  $f(R,\mathbf{T}^{2})=f_{1}(R)+f_{2}(R)f_{3}(\mathbf{T}^2)$}
Here we report the field equations for the second model.\\

\textbf{\textit{The (01) component is given by,}}
\begin{equation}\label{fieldeq01m2}
r\left[f_{1}(R)+f_{2}(R)f_{3}(T^{2})-2\rho\right]-\left(f_{1}'(R)
+f_{2}'(R)f_{3}(T^{2})\right)m''=0
\end{equation}

%$$16\left(f_{1}'''(R)+f_{2}'''(R)f_{3}(T^{2})\right)m'^{2}-4rm'\left[-3r\left(f_{1}''(R)
%+f_{2}''(R)f_{3}(T^{2})\right)+2f_{2}''(R)f_{3}'(T^{2})r^{2}\lambda\rho'\right.$$

%$$\left.+2\left(f_{1}'''(R)+f_{2}'''(R)f_{3}(T^{2})\right)\left(m''+rm'''\right)\right]
%+r^{2}\left[f_{2}'(R)f_{3}''(T^{2})r^{4}\lambda^{2}\rho'^{2}+\left(f_{1}'''(R)+f_{2}'''(R)f_{3}(T^{2})\right)m''^{2}\right.$$

%$$\left.+2rm''\left\{-3\left(f_{1}''(R)+f_{2}''(R)f_{3}(T^{2})\right)
%+\left(f_{1}'''(R)+f_{2}'''(R)f_{3}(T^{2})\right)m'''\right\}+2f_{2}''(R)f_{3}'(T^{2})r^{2}\lambda\rho'\left(m''+rm'''\right)\right.$$

%\begin{equation}\label{fieldeq11m2}
%\left.+r^{2}\left\{\left(f_{1}'''(R)+f_{2}'''(R)f_{3}(T^{2})\right)m'''^{2}
%+r\left(f_{2}'(R)f_{3}'(T^{2})r\lambda\rho''+\left(f_{1}''(R)+f_{2}''(R)f_{3}(T^{2})\right)m^{iv}\right)\right\}\right]=0
%\end{equation}

\textbf{\textit{The (22) and (33) components are given by,}}
\begin{equation}\label{fieldeq22m2}
\left(f_{1}'(R)+f_{2}'(R)f_{3}(T^{2})\right)m'-r^{2}\left\{\omega\rho+\frac{1}{2}
\left(f_{1}(R)+f_{2}(R)f_{3}(T^{2})\right)\right\}=0
\end{equation}

Here also the other components of the field equations are reported
in the appendix section of the paper.

%The (33) component is given by,
%\begin{equation}\label{fieldeq33m2}
%\left[\left(f_{1}'(R)+f_{2}'(R)f_{3}(T^{2})\right)m'-r^{2}\left\{\omega\rho+\frac{1}{2}\left(f_{1}(R)
%+f_{2}(R)f_{3}(T^2)\right)\right\}\right]\sin^{2}\theta=0
%\end{equation}

%%%%%%%%%%%%%%%%%%%%%%%%%%%%%%%%%%%%%
\subsection{Solution of the system}
%%%%%%%%%%%%%%%%%%%%%%%%%%%%%%%%%%%%%
In order to find solution for the field equations we need to
consider special toy-models as case studies. Here we can consider
various forms of functionalities for $f(R)$ and
$f(\mathbf{T}^{2})$ for both types of models.

\subsubsection{Model-1}

Here we consider the specific functionalities
\textbf{$f_{1}(R)=g_{1}R^{\beta_{1}},~~
f_{2}(\mathbf{T}^2)=g_{2}(\mathbf{T}^{2})^{\beta_{2}}$}, where
$g_{1}, \beta_{1}, g_{2}, \beta_{2}$ are constants. In this case
we have
$f(R,\mathbf{T}^2)=g_{1}R^{\beta_{1}}+g_{2}(\mathbf{T}^{2})^{\beta_{2}}$.
This model was studied in ref.\cite{rudra1}, where a dynamical
system analysis was performed. We recover GR for $g_{2}=0$ and
$g_{1}=\beta_{1}=1$. When $g_{1}=\beta_{1}=\beta_{2}=1$ then the
model reduces to the one used in ref.\cite{board1}. When
$g_{1}=\beta_{1}=1$ and $g_{2}\neq0, \beta_{2}\neq0$, then we
recover the Energy Momentum Powered Gravity (EMPG) \cite{board1}.
In this model $g_{1}$ and $g_{2}$ play the role of the coupling
parameters between the curvature and the matter sectors.
\\\\

Here the (01) component of the field equations give,
\begin{equation}\label{case1diff1}
4^{\beta_{2}}g_{2}r^{2}\left(n^{2}\omega^{2}m^{2}\right)^{\beta_{2}}+g_{1}\left(\frac{2m'+rm''}
{r^{2}}\right)^{\beta_{1}-1}\left(2m'-r\left(\beta_{1}-1\right)m''\right)-2nr^{2}m=0
\end{equation}

Solving the above equation we get for $\beta_{1}=\beta_{2}=1$,
\begin{equation}\label{solcase1diff1}
m(t,r)=\frac{e^{\frac{nr^{3}}{3g_{1}}}}{e^{h_{1}(t)}+2g_{2}n\omega^{2}e^{\frac{nr^{3}}{3g_{1}}}}
\end{equation}
where $h_{1}(t)$ is an arbitrary function of time $t$.

The (22) or (33) component becomes,
\begin{equation}\label{case1diff2}
r^{2}\left(2n\omega
m+4^{\beta_{2}}g_{2}\left(n^{2}\omega^{2}m^{2}\right)^{\beta_{2}}\right)
-g_{1}\left(2\left(\beta_{1}-1\right)m'-rm''\right)\left(\frac{2m'+rm''}{r^{2}}\right)^{\beta_{1}-1}=0
\end{equation}

Solving the above equation we get for $\beta_{1}=\beta_{2}=1$ and
for $\omega=0$ (dust)
\begin{equation}\label{solcase1diff2}
m(t,r)=h_{1}(t)+rh_{2}(t)
\end{equation}
where $h_{1}(t)$ and $h_{2}(t)$ are arbitrary functions of time
$t$. It is expected that the solution in eqn.(\ref{solcase1diff1})
will reduce to the solution given in eqn.(\ref{solcase1diff2}) for
some suitable initial conditions. So we will use the solution
given by eqn.(\ref{solcase1diff1}) for our collapse study because
of
its generic nature.\\\\

%\textbf{Case-2: $f_{1}(R)=R,~~
%f_{2}(T^2)=g_{2}\ln(\beta_{2} T^{2})$}, where $g_{2}, \beta_{2}$ are constants.\\

%In this case we have the model $f(R,T^2)=R+g_{2}\ln(\beta_{2}
%T^{2})$, $\beta_{2}>0$. This model is known as the
%Energy-Momentum-Log Gravity (EMLG) \cite{emlg1}, where $f(T^2)$ is
%given in the form of a logarithmic function of $T^2$. Obviously
%for $g_{2}=0$, we recover GR from this model.

\subsubsection{Model-2}

Here we consider the functional forms
\textbf{$f_{1}(R)=g_{1}R^{\beta_{1}},~~
f_{2}(R)=g_{2}R^{\beta_{2}},~~f_{3}(T^{2})=(T^2)^{\beta_{3}}$},~
where $g_{1}, \beta_{1}, g_{2}, \beta_{2}, \beta_{3}$ are
constants. So we have
$f(R,T^2)=g_{1}R^{\beta_{1}}+g_{2}R^{\beta_{2}}(T^{2})^{\beta_{3}}$.
Here also we recover GR for $g_{2}=0$ and $g_{1}=\beta_{1}=1$. For
$g_{1}=0$, we get the model studied in ref.\cite{rudra1}.

Here the (01) component of the field equations give,

$$-2nm\left(2m'+rm''\right)+g_{1}\left(\frac{2m'+rm''}{r^{2}}\right)^{\beta_{1}}
\left(2m'-r\left(\beta_{1}-1\right)m''\right)+g_{2}
\left(4n^{2}\omega^{2}m^{2}\right)^{\beta_{3}}\times$$

\begin{equation}\label{mod2diff1}
\left(\frac{2m'+rm''}{r^{2}}\right)^{\beta_{2}}\left(2m'-r\left(\beta_{2}-1\right)m''\right)=0
\end{equation}

Solving the above equation for $\beta_{1}=\beta_{2}=\beta_{3}=1$
we get,
\begin{equation}\label{solmod2diff1}
m(t,r)=\frac{\sqrt{g_{1}W\left[\frac{4g_{2}n^{2}\omega_{2}e^{\frac{2nr^{3}+6h_{1}(t)}{3g_{1}}}}{g_{1}}\right]}}{2\sqrt{g_{2}}n\omega}
\end{equation}
where $W[y]$ is the Lambert W function and $h_{1}(t)$ is an
arbitrary function of time.

The (22) or (33) component becomes,

$$2n\omega
m\left(2m'+rm''\right)-g_{1}\left(2\left(\beta_{1}-1\right)m'-rm''\right)
\left(\frac{2m'+rm''}{r^{2}}\right)^{\beta_{1}}-g_{2}\left(4n^{2}\omega^{2}m^{2}\right)^{\beta_{3}}\times$$

\begin{equation}\label{mod2diff2}
\left(2\left(\beta_{2}-1\right)m'-rm''\right)\left(\frac{2m'+rm''}{r^{2}}\right)^{\beta_{2}}=0
\end{equation}

Solving the above equation we get for $\omega=0$ (dust) and
$\beta_{1}=\beta_{2}=\beta_{3}=1$,
\begin{equation}\label{solmod2diff2}
m(t,r)=h_{1}(t)+rh_{2}(t)
\end{equation}
where $h_{1}(t)$ and $h_{2}(t)$ are arbitrary functions of time
$t$. Quite naturally due to its generic nature we will use the
solution given by eqn.(\ref{solmod2diff1}) for further analysis.

%%%%%%%%%%%%%%%%%%%%%%%%%%%%%%%%%
\section{Gravitational collapse: Nature of singularity}
%%%%%%%%%%%%%%%%%%%%%%%%%%%%%%%%%
An object of mass whose internal pressure is not enough to offset
the gravitational forces pulling it inward is said to be
undergoing gravitational collapse. A black hole's creation is the
most well-known result of gravitational collapse. The formation of
a big star marks the start of the process. A star's life is
characterized by nuclear fusion, which turns hydrogen into helium
and other heavier elements. The star reaches a critical point when
its nuclear fuel runs exhausted, at which time the equilibrium
between internal pressure pushing outward and gravitational forces
pulling inward becomes unstable. When the core of a large star can
no longer withstand the force of gravity, it collapses due to the
pull of gravity. This collapse can occur quickly, raising the
temperature and density. The collapsing core's density greatly
rises as it continues to fall. This results in the production of a
singularity, or a point with infinite density and curvature, in
the context of classical general relativity. Usually, this
singularity is concealed within an area known as the event
horizon. An event horizon forms around the singularity if there is
not enough angular momentum or charge in the collapsing item to
stop it. The event horizon is the point beyond which the
tremendous pull of gravity prevents anything from escaping, not
even light. This signifies the black hole's creation. In the
presence of density inhomogeneities an event horizon forms and a
black hole is created in a homogeneous collapse, in which the
collapsing object has a uniform density and spherical symmetry.
This process is well-described. The event horizon encloses the
singularity. The process of collapse gets more difficult in the
presence of asymmetries or density inhomogeneities. It is still
feasible for an event horizon to form, but the specifics will
depend on the mass and energy distribution. The Cosmic Censorship
Hypothesis raises the question of whether or not a naked
singularity (a singularity devoid of an event horizon) can form.
According to this theory, which was put forth by Roger Penrose
\cite{penrose}, singularities ought to always be concealed within
event horizons. If the theory is correct, then any collapsing
object would become a black hole with an event horizon, and
density inhomogeneities will not be able to resist its formation,
thus resulting in a naked singularity. A thorough understanding of
general relativity is necessary to comprehend the behavior of
gravitational collapse and the creation of horizons in the
presence of density inhomogeneities. This is still an area of
ongoing theoretical physics research. One important frontier in
our attempt to unify general relativity and quantum mechanics is
the nature of singularities and horizons, where the effects of
quantum gravity may become relevant.

Here we are concerned with the nature of singularity formed as a
result of the collapse. So the whole study revolves round the fact
that whether or not trapped surfaces are formed surrounding the
singularity. Now this is not so straightforward as it seems to be.
There are other essential ideas to consider over here. Not only
the formation of the horizons, but also the timing of the
formation of the horizons play crucial role in determining the
nature of the singularity formed. If the horizon forms far before
the formation of the singularity, then it is always censored from
the external observer, leading to a black hole. But if the
formation of the trapped surface is delayed, such that the
singularity forms first and then the trapped surface begins to
form, then the singularity is free to share information with the
external observer for some finite amount of time, thus being naked
\cite{collrev1}. It is found that inhomogeneities in the initial
density profile of the collapsing matter can result in the
postponement of the formation of the horizon \cite{collrev1}.
Moreover the presence of shear in the system can also hinder the
formation of the trapped surface for a finite amount of time
\cite{collrev1}. Since we have seen that EMSG has terms that mimic
some properties of loop quantum gravity, there may be some quantum
fluctuations at the fundamental level that can create some
inhomogeneities in the matter density profile. This can
substantially delay the formation of trapped surfaces if not
completely hinder its formation, thus making the singularity
naked. So it seems that the rate of collapse plays an important
role in determining the nature of the singularity formed. A slow
collapse will keep the density distribution more or less
homogeneous over time, thus resulting in a black hole, whereas a
fast collapse will tend to form a naked singularity due to the
inhomogeneity in matter density created over time. So we are
motivated to explore these effects for EMSG.

For any relativistic metric, we know that $ds^{2}=0$ gives the
directions along which the light rays travel. This is evident from
the study of light cones in the special theory of relativity. Here
we would probe the existence of outgoing light rays (radial null
geodesics) from the core of the central singularity. The equation
for such geodesics can be found from the metric (\ref{vaidya}) by
equating $ds^{2}=0$ and
$d\Omega_{2}^{2}=d\theta^{2}+\sin^{2}\theta d\phi^{2}=0$ so as to
get
\begin{equation}
\frac{dt}{dr}=\frac{2}{\left(1-\frac{m(t,r)}{r}\right)}
\end{equation}
This is a differential equation, which possess a singularity at
the point $r=0,~t=0$. There is a complete breakdown of
mathematical and physical structures at the singularity and hence
we need to examine the limiting behaviour of the trajectories as
one approaches the singularity. In order to conduct such a study,
we consider a parameter $X=t/r$. We will study the limiting
behaviour of the function $X$ as one proceeds towards the
singularity at $r=0,~t=0$ along the trajectory of the radial null
geodesic. Suppose the limiting value of $X$ at $r=0,~t=0$ is given
by $X_{0}$, then L'Hospital's rule yields
\begin{eqnarray}\label{X0}
\begin{array}{c}
X_{0}\\\\
{}
\end{array}
\begin{array}{c}
=\lim X \\
\begin{tiny}t\rightarrow 0\end{tiny}\\
\begin{tiny}r\rightarrow 0\end{tiny}
\end{array}
\begin{array}{c}
=\lim \frac{t}{r} \\
\begin{tiny}t\rightarrow 0\end{tiny}\\
\begin{tiny}r\rightarrow 0\end{tiny}
\end{array}
\begin{array}{c}
=\lim \frac{dt}{dr} \\
\begin{tiny}t\rightarrow 0\end{tiny}\\
\begin{tiny}r\rightarrow 0\end{tiny}
\end{array}
\begin{array}{c}
=\lim \frac{2}{\left(1-\frac{m(t,r)}{r}\right)} \\
\begin{tiny}t\rightarrow 0\end{tiny}~~~~~~~~~~~~\\
\begin{tiny}r\rightarrow 0\end{tiny}~~~~~~~~~~~~
 {}
\end{array}
\end{eqnarray}
Using the mass terms that we obtained previously the above limit
will provide an algebraic equation in terms of $X_{0}$. We will
hunt for the roots of this equation, which actually represent the
directions of the tangents to the outgoing geodesics. In principle
we should only be interested in the real roots of the equation,
since they represent the real exchange of information with the
external observers via escaping radiation. In our mathematical
set-up, any positive real root of the equation will indicate the
gradient of the tangent to an outgoing null geodesic. So it can be
concluded that the existence of real positive roots of the
obtained algebraic equation corresponds to the creation of a naked
singularity as a result of the stellar collapse.

When a single null geodesic get away the singularity, it
corresponds to a single wavefront being emanated from the central
singularity and illuminating the external observer. In such a
case, the singularity would be observable instantaneously to a
distant observer, before visibility is obstructed. This will
result in a \textit{locally naked singularity}. Physically
speaking, here the formation of the trapped surface is delayed for
a little amount of time, when the singularity becomes visible
temporarily. But this temporary exposure might not be adequate for
a total trade-off of data and information between the singularity
and the distant spectator. So for a comprehensive trade-off of
information, the singularity needs to be visible for an elongated
time interval. This is possible if a bundle of null geodesics
emanates from the central singularity, making it \textit{globally
naked}. Our theoretical set-up provides a perfect mechanism to
explore this quite easily by examining the count of real positive
roots derived from the algebraic equation. The above discussed
mathematical formulation for investigating the nature of
singularity formed as the final state of a stellar collapse was
first proposed by Joshi, Singh and Dwivedi in several of their
papers \cite{cch2, cch3, sing, ns5}. With the theoretical
framework ready, we will now proceed to examine the models
separately.

The physical characteristics of the collapsing object, the
equation of state of the matter involved, and the particulars of
the collapse process all influence the threshold mass at which
black hole formation would be prevented. It's crucial to remember
that the intricacy of the issue and the range of possibilities
that can result in the formation of a black hole make it difficult
to provide accurate quantitative estimations. However, a broad
rundown of a few important ideas can be provided. For a
non-rotating, non-charged star supported by electron degeneracy
pressure, the Chandrasekhar limit \cite{chandra1} is a critical
mass beyond which electron degeneracy pressure cannot
counterbalance gravity, leading to gravitational collapse. This
limit is approximately $1.44 M_\odot$, where $M_\odot$ is one
solar mass. The pressure of neutron degeneracy opposes the
collapse of more massive objects, like neutron stars. The total
object mass limit (Tolman-Oppenheimer-Volkof (TOV) limit)
\cite{tov1, tov2} is the maximum mass that a neutron star can have
before collapsing into a black hole. This limit is around $2.16
M_\odot$, although the exact number is unclear due to difficulties
in the equation of state for neutron-rich matter. A distinct
mechanism is involved in the case of big stars. Pair production
can result in a drastic drop in pressure for stars with masses
around $100-250 M_\odot$, which can trigger a pair-instability
supernova. As a result of the star's total disruption in these
situations, black hole creation is prevented. Under some
circumstances, black holes can emerge straight ahead of time
without the need for an intermediary stage for a supermassive
star. This may occur in areas with low metallicity and high
density where radiation pressure and nuclear burning do not impede
the collapse. Although less clearly defined, the mass threshold
for stars that collapse directly to black holes may be in the
range of $10^2-10^3 M_\odot$. It is crucial to stress that these
are only approximations and that the real mass thresholds would
vary depending on the particular circumstances and characteristics
of the collapsing object. Furthermore, we are constantly improving
our understanding of these processes through astrophysical
observations. The above quantitative estimates are provided
considering standard general relativity as the background theory
of gravity. In case of our study, since the correction terms of
EMSG possess quantum gravitational effects, they will try to
hinder the collapsing procedure thus trying to avoid the ultimate
singularity. So the intensity of the collapsing mechanism will be
reduced depending on several other factors. Therefore for EMSG
there is a possibility that each of the quantitative estimates
given above will be raised. For example for a non-rotating,
non-charged star the Chandrasekhar Limit may be raised from $1.44
M_\odot$. So the new mass limit $M_n$ at which the collapse occurs
will be $M_n>1.44 M_\odot$. Similar argument goes for the other
cases as well.

\subsection{Model-1: $f(R,\textbf{T}^{2})=R+\alpha \textbf{T}^2$}
Here we will study a particular toy model based on the model-1
that was discussed in the previous section. The model is given by
$f(R,\textbf{T}^{2})=R+\alpha \textbf{T}^2$, where $\alpha$ is the
coupling parameter. This model is widely studied in the literature
and is cosmologically a successful model. So here we are
interested in studying the model in a collapsing scenario. So this
model will be retrieved from model-1 for
$\beta_{1}=\beta_{2}=g_{1}=1$ and $g_{2}=\alpha$. Using
eqn.(\ref{solcase1diff1}) and eqn.(\ref{X0}) for this model we get
(on expanding the exponentials and taking only the linear terms),

\begin{eqnarray}\label{X01}
\frac{2}{X_{0}}=
\begin{array}llim\\
\begin{tiny}t\rightarrow 0\end{tiny}\\
\begin{tiny}r\rightarrow 0\end{tiny}
\end{array}
\left[1-\frac{1+\frac{nr^{3}}{3}}{r\left\{1+h_{1}(t)+2\alpha
n\omega^{2}\left(1+\frac{nr^{3}}{3}\right)\right\}}\right]
\end{eqnarray}
Now we will consider a self-similar collapsing scenario by the
choice of arbitrary function $h_{1}(t)=\xi_{1}t^{-1}$, where
$\xi_{1}$ is an arbitrary constant. Using this choice in
eqn.(\ref{X01}) we get the following algebraic equation in $X_{0}$
\begin{equation}\label{alg1}
X_{0}^{2}-\xi_{1}X_{0}+2\xi_{1}=0
\end{equation}
Solving the above equation we get the following two roots
\begin{equation}\label{root1}
X_{0_{1,2}}=\frac{1}{2}\left(\xi_{1}\pm
\sqrt{\xi_{1}}\sqrt{\left(\xi_{1}-8\right)}\right)
\end{equation}
Here we will consider the positive root as $X_{0_{1}}$ and the
negative root as $X_{0_{2}}$. To get a realistic scenario we
should have $\xi_{1}\leq0$ and $\xi_{1}\geq8$. Now depending on
the signature of the roots $X_{0_{1}}$ and $X_{0_{2}}$ the
collapse results in different outcomes. The conditions for getting
a locally naked singularity are ($X_{0_{1}}>0$~~ $\&$~~
$X_{0_{2}}<0$)~~OR~~($X_{0_{1}}<0$~~$\&$~~$X_{0_{2}}>0$). For a
globally naked singularity we should have~~ $X_{0_{1}}>0$~~
$\&$~~$X_{0_{2}}>0$. Finally for $X_{0_{1}}<0$~~
$\&$~~$X_{0_{2}}<0$~~ the collapse results in a black hole
according to our theoretical framework. To understand the trend of
the roots they are plotted in Fig.1 and Fig.2 respectively against
$\xi_{1}$. It clearly seen from the plots that the roots are not
realistic in the range $0<\xi<8$ as discussed before. Hence there
are no well-defined trajectories of the roots in that parameter
range. From Fig.1 it is seen that the $X_{0_{1}}<0$ for $\xi_1<0$
and $X_{0_{1}}>0$ for $\xi_1>8$. From Fig.2 it is evident that
$X_{0_{2}}>0$ throughout its domain of definition. There are two
disconnected trajectories (obvious due to the region $0<\xi_{1}<8$
being outside the domain of definition of the root) and they reach
asymptotic nature near $\xi_1=0$ and $\xi_1=8$. Now combining the
information obtained from both the figures we have $\Rightarrow$
$X_{0_{1}}>0$,~$X_{0_{2}}>0$ for $\xi_1>8$, \&
$X_{0_{1}}<0$,~$X_{0_{2}}>0$ for $\xi_1<0$. So we for $\xi_{1}>8$,
the collapse results in a globally naked singularity while for
$\xi_{1}<0$, a locally naked singularity forms. We further see
that formation of a black hole is not an option here. This is
probably because the EMSG model has terms which have reminiscence
of quantum gravity flavour, which mimic loop quantum gravity
\cite{emsgorg}. So these models have a tendency to bounce at a
finite time rather than end in a trapped singularity.

\begin{figure}\label{f1}
~~~~~~~~~\includegraphics[height=1.7in,width=2.5in]{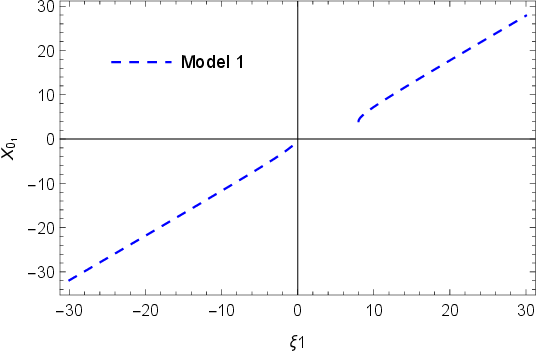}~~~~~~~~~~~~~~~~~~\includegraphics[height=1.7in,width=2.5in]{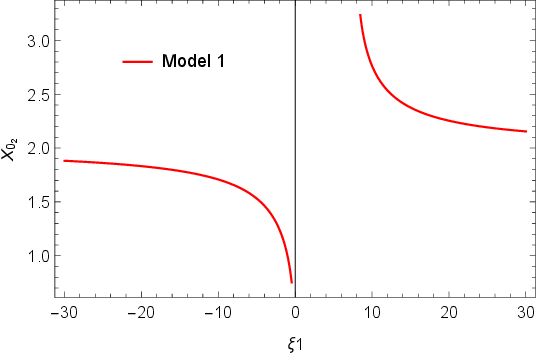}~~~~~~~\\

~~~~~~~~~~~~~~~~~~~~~~~~~~~~Fig.1~~~~~~~~~~~~~~~~~~~~~~~~~~~~~~~~~~~~~~~~~~~~~~~~~~~~~~~~~~~~~Fig.2~~~~~~~~~\\

\vspace{1mm} \textit{\textbf{Figs.1 and 2} show the variation of
the roots $X_{0_{1}}$ and $X_{0_{2}}$ respectively for different
values of $\xi_1$ for Model-1.}
\end{figure}

\subsection{Model-2: $f(R,\textbf{T}^{2})=f_{0}R\textbf{T}^2$}
In this case we intend to study a particular toymodel based on
model-2 discussed in the previous section. The model is given by
$f(R,\textbf{T}^{2})=f_{0}R\textbf{T}^2$, where $f_{0}$ is the
coupling parameter. Now using eqn.(\ref{fieldeq01m2}) and
(\ref{X0}) we get for this model,
\begin{eqnarray}\label{X02}
\frac{2}{X_{0}}=
\begin{array}llim\\
\begin{tiny}t\rightarrow 0\end{tiny}\\
\begin{tiny}r\rightarrow 0\end{tiny}
\end{array}
\left[1-\frac{1}{\omega
r}\sqrt{\frac{r^{3}+12f_{0}n\omega^{2}h_{2}(t)}{6f_{0}n}}\right]
\end{eqnarray}
where $h_{2}(t)$ is an arbitrary function of $t$. Now considering
self similar collapsing scenario, we choose
$h_{2}(t)=\xi_{2}t^{2}$, where $\xi_{2}$ is an arbitrary constant.
Using this choice in eqn.(\ref{X02}) we get the following
algebraic equation in $X_{0}$
\begin{equation}\label{alg2}
\sqrt{2\xi_{2}}~X_{0}^{2}-X_{0}+2=0
\end{equation}
Solving the above equation we get the roots as,
\begin{equation}\label{root2}
X_{0_{1,2}}=\frac{4}{1\pm\sqrt{1-8\sqrt{2\xi_{2}}}}
\end{equation}
The most striking feature of these roots are the extreme level of
constraint that they suffer from in order to take real values. The
range in which these root actually take some realistic values are
as narrow as $0<\xi_{2}<0.0078$. So we see that this model is
highly delicate and requires a considerable amount of fine tuning
to derive meaningful result out of it. To gain insights into the
trend of the roots, they are plotted in Fig.3 and Fig.4
respectively against $\xi_{2}$. From the figures it is evident
that in the domain of definition, both the roots remain in the
positive level. This shows that we have a globally naked
singularity as the outcome of collapse. The quantum gravity
effects coming into play restricts the singularity from being
trapped inside the horizon.

\begin{figure}\label{f1}
~~~~~~~~~\includegraphics[height=1.7in,width=2.5in]{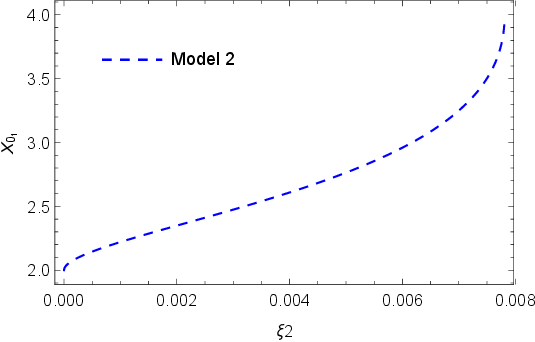}~~~~~~~~~~~~~~~~~~\includegraphics[height=1.7in,width=2.5in]{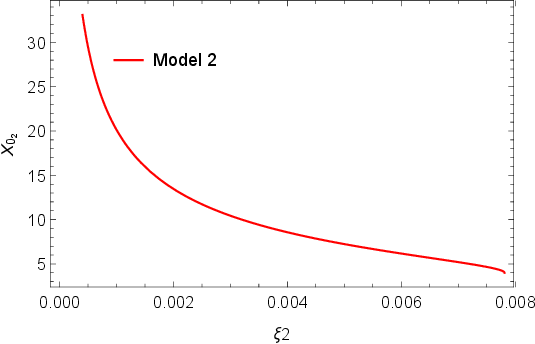}~~~~~~~\\

~~~~~~~~~~~~~~~~~~~~~~~~~~~~Fig.3~~~~~~~~~~~~~~~~~~~~~~~~~~~~~~~~~~~~~~~~~~~~~~~~~~~~~~~~~~~~~Fig.4~~~~~~~~~\\

\vspace{1mm} \textit{\textbf{Figs.3 and 4} show the variation of
the roots $X_{0_{1}}$ and $X_{0_{2}}$ respectively for different
values of $\xi_2$ for Model-2.}
\end{figure}

%%%%%%%%%%%%%%%%%%%%%%%%%%%%%%%%%%%%%%%%%%%%%%%%%%%%%%%%%%%%%%%%%%%%%
\subsection{Strength of the singularity (Curvature growth near the
singularity)} %%%%%%%%%%%%%%%%%%%%%%%%%%%%%%%%%%%%%%%%%%%%%%%%%%%%%%

The amount of curvature inflicted on the surrounding spacetime
determines the gravitational strength of a singularity. It is the
estimate of the destructive capacity of the singularity on any
matter that passes by it. Most theories of gravity have been
crippled by the presence of singularities in their theoretical
set-up. Various theoretical mechanisms of elimination of such
singularities have been proposed in literature from time to time,
but they are substantially alien in nature and far from being
acceptable. We know that singularities are depressions in the
fabric of the otherwise continuous and smooth spacetime. In case
of a weak singularity the crater is shallow and a continuous
extension of spacetime is feasible right through the singularity.
Mathematically this idea is analogous to a removable
discontinuity, which can rehabilitate the discontinuity of
spacetime at a singularity. From the above discussion we see that
there is enough motivation in investigating whether a singularity
is strong or weak in nature. According to Tipler \cite{Tipler} a
curvature singularity will be strong if any object coming in its
vicinity or colliding with it is squeezed to zero volume. In
Ref.\cite{Tipler} the condition for a strong singularity is given
as,
\begin{eqnarray}\label{tipu}
\begin{array}{c}
S=\lim \tau^{2}\psi \\
\begin{tiny}\tau\rightarrow 0\end{tiny}\\
\end{array}
\begin{array}{c}
=\lim \tau^{2}R_{\mu\nu}K^{\mu}K^{\nu}>0 \\
\begin{tiny}\tau\rightarrow 0\end{tiny}\\
\end{array}
\end{eqnarray}
where $R_{\mu\nu}$ is the Ricci tensor, $\psi$ is a scalar given
by the relation $\psi=R_{\mu\nu}K^{\mu}K^{\nu}$, where
$K^{\mu}=dx^{\mu}/d\tau$ represents the tangent to the non
spacelike geodesics at the singularity and $\tau$ is the affine
parameter. In Ref.\cite{Maharaj} Mkenyeleye et al. have shown
that,
\begin{eqnarray}\label{maha}
\begin{array}{c}
S=\lim \tau^{2}\psi \\
\begin{tiny}\tau\rightarrow 0\end{tiny}\\
\end{array}
\begin{array}{c}\label{stren}
=\frac{1}{4}X_{0}^{2}\left(2\dot{m_{0}}\right) \\
\begin{tiny}~\end{tiny}\\
\end{array}
\end{eqnarray}
where
\begin{eqnarray}
\begin{array}{c}
m_{0}=\lim~ m(t,r) \\
\begin{tiny}t\rightarrow 0\end{tiny}\\
\begin{tiny}r\rightarrow 0\end{tiny}
\end{array}
\end{eqnarray}
and
\begin{eqnarray}\label{massd}
\begin{array}{c}
\dot{m_{0}}=\lim \frac{\partial}{\partial~t}\left(m(t,r)\right) \\
\begin{tiny}t\rightarrow 0\end{tiny}\\
\begin{tiny}r\rightarrow 0\end{tiny}
\end{array}
\end{eqnarray}
In ref. \cite{Maharaj} it has also been shown that the relation
between $X_{0}$ and the limiting values of mass is given by,
\begin{equation}\label{xmass}
X_{0}=\frac{2}{1-2m_{0}'-2\dot{m_{0}}X_{0}}
\end{equation}
where
\begin{eqnarray}\label{dashedmass}
\begin{array}{c}
m_{0}'=\lim \frac{\partial}{\partial~r}\left(m(t,r)\right) \\
\begin{tiny}t\rightarrow 0\end{tiny}\\
\begin{tiny}r\rightarrow 0\end{tiny}
\end{array}
\end{eqnarray}
and $\dot{m_{0}}$ is given by the eqn.(\ref{massd}).

Studies by Dwivedi and Joshi in Refs.\cite{strongcurvature, cch2}
revealed that any classical singularity in Vaidya spacetime in the
background of Einstein gravity should to be a very strong
curvature singularity going by the definition of Tipler. In
addition to this, the authors have also shown that the hypothesis
\cite{tiplernew} that the strong curvature singularities are
always trapped by horizons is not always true. It is expected that
in the background of the quantum fluctuations of
$f(R,\mathbf{T^2})$ gravity the strength of the singularity will
be considerably weakened if not totally eliminated. The structure
of such a naked singularity was explored in
Ref.\cite{strongcurvature1} and it was shown that the singularity
attains a directional nature as the curvature grows along the
geodesics, finally concluding in the singularity. Moreover for a
quantum regime it was evident that the singularity is
gravitationally weak in nature, which permitted a continuous
extension of the spacetime through the singularity
\cite{weakness}. Now we will study the strength of the
singularities for the respective models.

\subsection{Model-1: $f(R,\textbf{T}^{2})=R+\alpha \textbf{T}^2$}
For this model the mass parameter is calculated as
\begin{equation}\label{solcase1strength}
m(t,r)=\frac{e^{\frac{nr^{3}}{3}}}{e^{h_{1}(t)}+2\alpha
n\omega^{2}e^{\frac{nr^{3}}{3}}}
\end{equation}
Firstly we take the time derivative of the above expression to get
$\frac{\partial}{\partial~t}\left(m(t,r)\right)=\frac{\xi_{1}e^{\frac{nr^{3}}{3}+\frac{\xi_{1}}{t}}}{t^{2}\left(e^{\frac{\xi_{1}}{t}}
+2n\alpha\omega^{2}e^{\frac{nr^{3}}{3}}\right)^2}$. Taking the
limit of this as $r\rightarrow 0$ we get
$\frac{\xi_{1}e^{\frac{\xi_{1}}{t}}}{t^{2}\left(e^{\frac{\xi_{1}}{t}}
+2n\alpha\omega^{2}\right)^2}$. Now taking the limit of this
expression as $t\rightarrow 0$ we get an indeterminate form, which
is not removable by L'Hospital's method. To resolve this issue, we
resort to Taylor expansion of the exponential terms of mass given
in eqn.(\ref{solcase1strength}). Expanding and considering the
linear terms only we get
$m(t,r)=\frac{1+\frac{nr^{3}}{3}}{1+\frac{\xi_{1}}{t}+2\alpha
n\omega^{2}\left(1+\frac{nr^{3}}{3}\right)}$. It should be
mentioned here that the qualitative features of the collapse
mechanism are not hampered due to truncation of the series to
linear terms only. This is because we are going to take limits as
$r\rightarrow 0$ and $t\rightarrow 0$ to track the singularity,
and all the non-linear terms will behave exactly as the linear
term in the limiting scenario for mass. Taking the time derivative
for the above Taylor expanded expression for mass we get
$\frac{\left(1+\frac{nr^{3}}{3}\right)\xi_{1}}{t^{2}\left(1+\frac{\xi_{1}}{t}+2\alpha
n\omega^{2}\left(1+\frac{nr^{3}}{3}\right)\right)^{2}}$. Now
taking limit as $r\rightarrow0$ we get
$\frac{\xi_{1}}{t^{2}\left(1+\frac{\xi_{1}}{t}+2\alpha
n\omega^{2}\right)^{2}}$. We take the limit $t\rightarrow0$ of the
obtained expression and see that we have an indeterminate form.
But this time the indeterminacy can be removed by using
L'Hospital's rule and we get $\dot{m}_{0}=0$. Using this in eqn.
(\ref{maha}) we get,
\begin{equation}
S=\frac{1}{4}X_{0}^{2}\left(2\dot{m_{0}}\right)=0
\end{equation}
Therefore according to the definition of the strength of a
singularity by Tipler given in eqn.(\ref{tipu}), we have a
gravitationally weak singularity for this model. This is a direct
result of the quantum gravity terms of the theory, that eases the
singularity and gives us a less severe pathology to handle.

\subsection{Model-2: $f(R,\textbf{T}^{2})=f_{0}R\textbf{T}^2$}
For this model the mass parameter is given by,
\begin{equation}\label{solcase2strength}
m(t,r)=\frac{1}{\omega
}\sqrt{\frac{r^{3}+12f_{0}n\omega^{2}\xi_{2}t^{2}}{6f_{0}n}}
\end{equation}
Taking time derivative of the above expression we get
$\frac{2\xi_{2}\omega
t}{\sqrt{\frac{r^{3}}{6f_{0}n}+2\xi_{2}\omega^{2}t^{2}}}$. Now
taking limit as $r\rightarrow 0$ yields
$\frac{\sqrt{2}\xi_{2}\omega t}{\sqrt{\xi_{2}\omega^{2}t^{2}}}$.
Considering $\xi_{2}>0$, in the above expression we find the
$t\rightarrow 0$ limit as $\dot{m}_{0}=\sqrt{2\xi_{2}}$. For
$\xi_{2}<0$ we do not have real values. So using the result in
eqn. (\ref{maha}) we get,
\begin{equation}
S=\frac{1}{4}X_{0}^{2}\left(2\dot{m_{0}}\right)=\frac{1}{4}X_{0}^{2}\left(2\sqrt{2\xi_{2}}\right),~~~~~for
~~\xi_{2}>0
\end{equation}
Now the sign of $S$ completely depends upon whether we take the
positive or the negative square root of $\sqrt{\xi_{2}}$. If we
take the positive square root, $S>0$, and vice versa. So we see
that in this theoretical framework there is provision for both
strong and weak singularities depending on the initial conditions.
Here we see that the quantum gravitational effects of the model is
subdued compared to model-1. This results in both strong and weak
singularities depending on the initial conditions. So the nature
of coupling between the matter and the geometry sectors are
playing crucial roles in determining the nature of singularity
formed as a result of the collapse.

%%%%%%%%%%%%%%%%%%%%%%%%%%%%%%%%%%%%%%%%
\section{Conclusion and discussion}
%%%%%%%%%%%%%%%%%%%%%%%%%%%%%%%%%%%%%%%%
In this work, we have explored a collapsing scenario in the
background of $f(R,T_{\mu\nu}T^{\mu\nu})$ gravity. Since the
correction terms of these models are reminiscent of those coming
from the quantum gravity effects in loop quantum gravity or
braneworld models, the early radiation dominated universe is not
supposed to start from a singularity. Motivated from this we
wanted to investigate the effect of the quantum geometry on the
collapsing mechanism of a star in the background of this gravity.
The basic idea is to probe the possibility of a singularity-free
collapse. Even if a singularity forms, we wanted to find out
whether there is any possibility of formation of trapped surfaces
as the outcome of collapse. We modelled the spacetime of a massive
star by the Vaidya metric. The reason behind this choice is that,
the time dependent Vaidya metric represents the spacetime of a
realistic star. Moreover a collapsing procedure being a
time-dependent phenomenon is ideally suited for this metric.

The field equations were found and the mass term was calculated
from them. Now using this mass term we devised a theoretical
framework of the collapsing mechanism. The idea was to probe the
existence of outgoing radial null geodesics emanating from the
central singularity. This was obtained by considering the line
element $ds^{2}=0$ and $d\Omega_{2}^{2}=d\theta^{2}+\sin^{2}\theta
d\phi^{2}=0$. In the process we reached a differential equation in
$t$ and $r$. To explore the effects near the singularity, we
considered a limiting condition as $t\rightarrow0$ and
$r\rightarrow0$. This led us to an algebraic equation in terms of
a collapsing parameter $X_{0}$, whose signature actually
determined the nature of the singularity formed. A positive value
of $X_{0}$ indicated the presence of outgoing radial null
geodesics and hence showed the absence of trapped surfaces, i.e.
black holes. This would indicate that either there is no
singularity formed or even if there is a singularity it naked in
nature. If $X_{0}$ assumed negative values, it showed the absence
of any outgoing null geodesics. This indicated the probable
formation of event horizon, which meant the existence of a black
hole. So the whole mechanism was based on the nature of roots
obtained from our algebraic equation.

We applied this collapsing mechanism on two specific models of
$f(R,\textit{T}^{2})$ gravity with different types of couplings.
In the first model, we studied the functional form
$f(R,\textbf{T}^{2})=R+\alpha \textbf{T}^2$. It was found that
there was no possibility of the formation of a black hole. Either
the collapse did not result in a singularity or if it did so it
was a very weak naked singularity. Depending on the initial
conditions the naked singularity can be local or global. We think
that this can be totally attributed to the quantum geometry
effects coming from the correction terms. The quantum effects as
expected have either totally cured the singularity or even if it
is present, it does not have the ability to form trapped surfaces,
due to the loss of its extreme curvature effects (strength). In
the second model, we probed the functional form
$f(R,\textbf{T}^{2})=f_{0}R\textbf{T}^2$. It can be clearly seen
that the coupling effect between $R$ and $\textbf{T}^2$ is quite
different here in contrast to the first model. From the collapse
analysis it was seen that for this model also formation of trapped
surfaces is hindered, by the quantum gravity effects. The naked
singularity formed for this model was globally naked irrespective
of the initial conditions. However in the strength analysis it was
observed that the singularity may be weak or strong depending on
the initial conditions. This shows that the nature of coupling
between matter and geometry plays an important role in determining
the outcome of collapse. For minimal coupling the quantum gravity
effects are far more dominant compared to the non-minimal
coupling. This is the reason why the singularity considerably
eases out (weakens) for minimal coupling, while for the
non-minimal case there is a possibility for the formation of both
strong and weak singularities depending on the initial conditions.

If naked singularities exist, they would not have an event horizon
to conceal them from view, unlike black holes, which are
distinguished by this feature. Finding visible indicators to
differentiate between the two is a difficult endeavor since it
requires going beyond what is currently known about physics and
observational skills. However some observable indicators could be
useful in differentiating between naked singularities and black
holes. The Presence or absence of event horizon is the easiest way
to distinguish between naked singularities and black holes. There
could be differences in attributes between the accretion disks
surrounding naked singularities and black holes. For instance, in
the case of naked singularities, the structure and properties of
the accretion disk may be affected by the lack of an event
horizon. A system with a bare singularity may produce different
gravitational waves than a system with a black hole. Finding these
variations via gravitational wave measurements could yield
insightful information. A naked singularity may not have the same
gravitational pull on surrounding stuff as a black hole. Tracking
the movement of adjacent objects and the surrounding area may show
interesting patterns. Unlike black holes, bare singularities may
not produce Hawking radiation. A closer look at the radiation
characteristics close to the proposed singularity may reveal
further information about its existence. Analyzing quantum effects
close to the singularity may help distinguish bare singularities
from black holes. This would necessitate a greater comprehension
of quantum gravity, a notion that physicists are presently unable
to grasp. In contrast to events around black holes, events near
naked singularities may be more chaotic or display surprising
behaviors if they defy the Cosmic Censorship Hypothesis. It is
noteworthy that the aforementioned recommendations are theoretical
in nature and that there is currently a lack of a solid
theoretical foundation for naked singularities. Furthermore, it's
possible that our existing observational capabilities won't be
enough to find and identify these unusual things. Such severe
astrophysical phenomena demand a greater understanding of the
interactions between quantum mechanics and general relativity, an
area of ongoing theoretical physics research.

From the above discussion, we see that energy momentum squared
gravity can be a cure for the singularity problem existing in the
literature of cosmology. The asymptotic isolation of the
singularity for a distant observer is significantly eased out in
this case due to the presence of quantum fluctuations at the
fundamental level. Although in the early stages of research, but
still it has shown enough promise to be a potential candidate with
its quantum geometric correction terms which mimic the loop
quantum gravity or braneworlds. From the perspective of the cosmic
censorship hypothesis, this work seems to be a substantive
development. Hereby we reserve ourselves and do not go on to
declare the complete failure of cosmic censorship hypothesis, but
this work should be considered as more than a significant
counterexample of the hypothesis.

%%%%%%%%%%%%%%%%%%%%%%%%%%
\section*{Acknowledgments}
%%%%%%%%%%%%%%%%%%%%%%%%%%

The author acknowledges the Inter University Centre for Astronomy
and Astrophysics (IUCAA), Pune, India for granting visiting
associateship. The author also thanks the anonymous referee for
his/her valuable comments that helped to improve the quality of
the manuscript.

%%%%%%%%%%%%%%%%%%%%%%%%%%%%%%%%%%%%%%%
\section*{Data Availability Statement}
%%%%%%%%%%%%%%%%%%%%%%%%%%%%%%%%%%%%%%%
There is no data to report for this paper. No data has been used
to prepare this manuscript.

%%%%%%%%%%%%%%%%%%%%%%%%%%%%%%%%
\section*{Conflict of Interest}
%%%%%%%%%%%%%%%%%%%%%%%%%%%%%%%%
There are no known conflict of interest for the publication of
this paper.

%%%%%%%%%%%%%%%%%%%%%%
\section{Appendix}
%%%%%%%%%%%%%%%%%%%%%%
We present here the (00) and (11) component of the field
equations.

\subsection{Model-1}
\textbf{\textit{The (00) component of the field equation is given
by,}}

$$r^{4}\left\{-2f''(T^2)\rho^{2}\left(1+\omega\right)\left(1+3\omega\right)-2\sigma+2\rho
\left(\frac{m}{r}-1\right)\right\}+r^{3}\left(f_{1}(R)+f_{2}(T^2)\right)\left(r-m\right)$$

\begin{equation}\label{fieldeq00}
+r^{2}\left[f_{1}'(R)\left\{\left(m-r\right)m''+2\dot{m}\right\}
-2f_{1}''(R)\left(2\ddot{m}'+r\ddot{m}''\right)\right]-2f_{1}'''(R)\left(2\dot{m}'
+r\dot{m}''\right)^2=0
\end{equation}

\textbf{\textit{The (11) component is given by,}}
\begin{equation}\label{fieldeq11}
f_{1}'''(R)\left\{-4m'+r\left(m''+rm'''\right)\right\}^{2}+r^{2}f_{1}''(R)
\left(12m'-6rm''+r^{3}m^{(4)}\right)=0
\end{equation}

\subsection{Model-2}

\textbf{\textit{The (00) component is given by,}}

$$r^{4}\left[-2\sigma-2f_{2}(R)f_{3}'(T^{2})\rho^{2}\left(1+\omega\right)
\left(1+3\omega\right)+2\rho\left(\frac{m}{r}-1\right)-8f_{2}'(R)f_{3}''(T^{2})
\left(\omega-1\right)^{2}\dot{\rho}^{2}-4f_{2}'(R)f_{3}'(T^{2})\left(\omega-1\right)\ddot{\rho}\right]$$
$$+r^{3}\left(r-m\right)\left[f_{1}(R)+f_{2}(R)f_{3}(T^{2})\right]+r^{2}
\left[\left(f_{1}'(R)+f_{2}'(R)f_{3}(T^{2})\right)\left\{m''\left(m-r\right)
+2\dot{m}\right\}-8f_{2}''(R)f_{3}'(T^{2})\left(\omega-1\right)\dot{\rho}\times\right.$$
\begin{equation}\label{fieldeq00m2}
\left.\left(2\dot{m}'+r\dot{m}''\right)-2\left(f_{1}''(R)+f_{2}''(R)f_{3}(T^{2})\right)\left(2\ddot{m}'+r\ddot{m}''\right)\right]
-2\left[\left(2\dot{m}'+r\dot{m}''\right)^{2}\left(f_{1}'''(R)+f_{2}'''(R)f_{3}(T^{2})
\right)\right]=0
\end{equation}

\textbf{\textit{The (11) component is given by,}}

$$2r^{6}\left[2f_{2}'(R)f_{3}''(T^2)\left(\omega-1\right)^{2}\rho'^{2}+f_{2}'(R)f_{3}'(T^2)
\left(\omega-1\right)\rho''\right]+4r^{3}f_{2}''(R)f_{3}'(T^2)\left(\omega-1\right)\rho'
\left(-4m'+r\left(m''+rm'''\right)\right)$$
\begin{equation}\label{fieldeq11m2}
+r^{2}\left(12m'-6rm''+r^{3}m^{(4)}\right)\left(f_{1}''(R)+f_{2}''(R)f_{3}(T^2)\right)
+\left[-4m'+r\left(m''+rm'''\right)\right]^{2}\left(f_{1}'''(R)+f_{2}'''(R)f_{3}(T^2)\right)=0
\end{equation}

%%%%%%%%%%%%%%%%%%%%%%%%%%%%%%%%%%%%%%%%%%%%%%%%%%%%%%%%%%%%%%%%%%%%%
%%%%%%%%%%%%%%%%%%%%%%%%%%%%%%%%%%%%%%%%%%%%%%%%%%%%%%%%%%%%%%%%%5

\end{document}